\documentclass[11pt]{article}  

\usepackage{setspace}
\usepackage[]{graphicx}
\usepackage{amsmath}
\usepackage{amssymb}
\usepackage{subfigure}
\usepackage{epsfig}
\usepackage{wrapfig}
\usepackage{color}
\usepackage{rotating}
\usepackage{float}
 \usepackage{upgreek}
\usepackage{bm}
\usepackage{cite}

\DeclareGraphicsExtensions{.eps,.ps,.eps.gz,.ps.gz,.eps.Z,.EPS,.pdf,.PDF,.pstex,.png}

 \advance \textwidth by 0.1cm
 \advance \textheight by 0.06cm
\def\##1{{\bf #1}}
\def\=#1{\underline{\underline #1}}
\def\4#1{\underline{\underline{\underline{\underline #1}}}}

\def\.{\mbox{ \tiny{$^\bullet$} }}

\def\r#1{(\ref{#1})}

\def\t#1{\mbox{\tiny{#1}}}

\def\le{\left(}
\def\ri{\right)}
\def\les{\left[}
\def\ris{\right]}
\def\lec{\left\{}
\def\ric{\right\}}

\def\eps{\varepsilon}
\def\epso{\eps_{\scriptscriptstyle 0}}
\def\lambdao{\lambda_{\scriptscriptstyle 0}}
\def\muo{\mu_{\scriptscriptstyle 0}}
\def\ko{k_{\scriptscriptstyle 0}}
\def\etao{\eta_{\scriptscriptstyle 0}}
\def\co{c_{\scriptscriptstyle 0}}

\def\thetao{\theta_{\scriptscriptstyle 0}}
\def\psio{\psi_{\scriptscriptstyle 0}}

\def\bGamma{{\boldsymbol\Gamma}}
\def\Gammax{\Gamma_{\rm x}}
\def\Gammay{\Gamma_{\rm y}}
\def\Gammaz{\Gamma_{\rm z}}

\def\bsigma{{\boldsymbol\sigma}}
\def\btau{{\boldsymbol\tau}}

\def\epsr{\eps_r}
\def\mur{\mu_r}
\def\etar{\eta_r}

\def\sp{\#s}
\def\pinc{{\#p}_+}
\def\pref{{\#p}_-}

\def\inc{_{\rm inc}}
\def\refl{_{\rm refl}}
\def\refr{_{\rm refr}}

\def\Einc{{\#E}_{\rm inc}({\#r})}
\def\Erefl{{\#E}_{\rm refl}({\#r})}
\def\Erefr{{\#E}_{\rm refr}({\#r})}
\def\tErefr{\tilde{\#E}_{\rm refr}({\#r})}

\def\Etr{{\#E}_{\rm tr}({\#r})}

\def\Hinc{{\#H}_{\rm inc}({\#r})}
\def\Hrefl{{\#H}_{\rm refl}({\#r})}
\def\Hrefr{{\#H}_{\rm refr}({\#r})}
\def\tHrefr{\tilde{\#H}_{\rm refr}({\#r})}

\def\Htr{{\#H}_{\rm tr}({\#r})}

\def\spint{\bsigma}
\def\ppint{\btau_{+}}

\def\qx{q_{\rm x}}
\def\qy{q_{\rm y}}

\def\cpsio{\cos\psio}
\def\spsio{\sin\psio}
\def\cthetao{\cos\thetao}
\def\sthetao{\sin\thetao}

\def\as{a_{\rm s}}
\def\ap{a_{\rm p}}
\def\rs{r_{\rm s}}
\def\rp{r_{\rm p}}
\def\ts{t_{\rm s}}
\def\tp{t_{\rm p}}
\def\rhos{\rho_{\rm s}}
\def\rhop{\rho_{\rm p}}

\def\rss{r_{\rm ss}}
\def\rsp{r_{\rm sp}}
\def\rps{r_{\rm ps}}
\def\rpp{r_{\rm pp}}
\def\rhoss{\rho_{\rm ss}}
\def\rhosp{\rho_{\rm sp}}
\def\rhops{\rho_{\rm ps}}
\def\rhopp{\rho_{\rm pp}}
\def\tss{t_{\rm ss}}
\def\tsp{t_{\rm sp}}
\def\tps{t_{\rm ps}}
\def\tpp{t_{\rm pp}}

\def\Rss{R_{\rm ss}}
\def\Rsp{R_{\rm sp}}
\def\Rps{R_{\rm ps}}
\def\Rpp{R_{\rm pp}}
\def\Tss{T_{\rm ss}}
\def\Tsp{T_{\rm sp}}
\def\Tps{T_{\rm ps}}
\def\Tpp{T_{\rm pp}}

\def\ux{\hat{\#x}}
\def\uy{\hat{\#y}}
\def\uz{\hat{\#z}}

\def\ua{\hat{\#a}}
\def\ub{\hat{\#b}}
\def\uc{\hat{\#c}}

\def\dprime{{\prime\prime}}

\begin{document}

\begin{center}

\textbf{Planewave Response of a Simple Lorentz-Nonreciprocal Medium with Magnetoelectric Gyrotropy}\\

\textit{Akhlesh Lakhtakia  }

{The Pennsylvania State University, Department of Engineering Science and Mechanics,  University Park, PA 16802, USA}\\
{akhlesh@psu.edu}

\end{center}


\begin{abstract}
The simple Lorentz-nonreciprocal medium described by the constitutive relations
$\#D=\epso\epsr\#E-\bGamma\times\#H$ and
$\#B=\muo\mur\#H-\bGamma\times\#E$ is inspired by a specific spacetime metric,
$\bGamma$ being the magnetoelectric-gyrotopy vector.
Field representations in this medium can be obtained from those for
the isotropic dielectric-magnetic medium. When a plane wave is incident on a half space occupied
by the  Lorentz-nonreciprocal medium with magnetoelectric gyrotopy, theory shows that
the transverse component of the magnetoelectric-gyrotopy vector is responsible
for a rotation about the normal axis; furthermore, left/right reflection
asymmetry is exhibited. 
Additionally,
left/right transmission asymmetry is  exhibited by a planar slab composed of the  Lorentz-nonreciprocal medium with magnetoelectric gyrotopy.
The left/right asymmetries are of interest for
one-way  devices.

\end{abstract}

\section{Introduction}\label{sec:intro}

According to the Pl\'ebanski theorem \cite{Plebanski,deFelice},  relativistic
spacetime can be replaced by a bianisotropic continuum with constitutive relations
\begin{equation}
\label{conrel0}
\left.\begin{array}{l}
{\#D} =\epso\=\gamma\.{\#E} -\left({\bGamma}\times\=I\right)\. {\#H}
\\[5pt]
{\#B} =\muo\=\gamma\.{\#H} + \left({\bGamma}\times\=I\right)\.{\#E}
\end{array}\right\}\,,
\end{equation}
where $\=I$ is the identity dyadic \cite{Chen}, the dyadic $\=\gamma$  and
the vector $\#\Gamma$ emerge from the gravitational metric, $\muo$ is  the permeability of
gravitationally unaffected free space, and  $\epso$ is  the  permittivity of
gravitationally unaffected free space.
Several
composite materials have been theoretically formulated  \cite{Li1,ML1,Li3,ML3,Khvesh,Isabel,Pires}
to simulate certain characteristics of relativistic spacetime.

Whereas the dyadic $\=\gamma$ is real symmetric, the non-zero magnitude $\Gamma=\vert\bGamma\vert$ of
the   magneto\-electric-gyrotropy vector $\bGamma$ indicates that bianiostropic continuum is
nonreciprocal in the Lorentz sense \cite{Lorentz,Krowne} because the dyadic $\bGamma\times\=I$
is antisymmetric.  
For certain spacetime metrics, $\gamma$ is a diagonal dyadic, which  inspires a
medium with constitutive relations \cite{Jafri}
\begin{equation}
\label{conrels-EH}
\left.\begin{array}{l}
{\#D} =\epso\epsr{\#E} -{\bGamma}\times  {\#H}
\\[5pt]
{\#B} =\muo\mur{\#H} + {\bGamma}\times {\#E}
\end{array}\right\}\,.
\end{equation}
The electromagnetic response
characteristics of this medium, with isotropic relative permittivity $\epsr$ and relative permeability
$\mur$ but with anisotropic magnetoelectric properties, are the subject of this theoretical paper.

Section~\ref{rect} provides solutions for rectilinear
propagation in the Lorentz-nonreciprocal medium with magnetoelectric gyrotropy. Section~\ref{randr} is devoted to reflection of a plane wave
by a half space occupied by this medium, and Sec.~\ref{randt} to reflection and transmission
of a plane wave by a planar slab composed of the same medium.
The speed of light in free space, the free-space wavenumber, the
free-space wavelength, and the intrinsic impedance of free space are denoted by 
$\co=1/\sqrt{\epso\muo}$,
$\ko=\omega/\co$,
$\lambdao=2\pi/\ko$, and
$\etao=\sqrt{\muo/\epso}$, respectively, with $\omega$ as the angular frequency.
Vectors are in boldface, with Cartesian unit vectors  
identified as $\ux$, $\uy$ and $\uz$. Dyadics are double underlined. Matrixes are double underlined and enclosed
in square brackets. Complex quantities are exemplified by $\zeta=\zeta^\prime
+i\zeta^\dprime$, where $\zeta^\prime={\rm Re}(\zeta)$ and $\zeta^\dprime={\rm Im}(\zeta)$.

\section{Rectilinear propagation}\label{rect}

Electromagnetic field phasors in a medium described by Eqs.~\r{conrels-EH} can be obtained
after defining the   auxiliary field phasors \cite{LW1997}
\begin{equation}
\left.\begin{array}{l}
\tilde{\#E}(\#r) =\#E(\#r) \exp\left(-i\omega\bGamma\.\#r\right) 
\\[5pt]
\tilde{\#H}(\#r) =\#H(\#r) \exp\left(-i\omega\bGamma\.\#r\right)
\end{array}
\right\}\,. 
\label{ehint-def}
\end{equation}
These auxiliary field phasors are governed by the differential equations
\begin{equation}
\label{diffeq-ehint}
\left.\begin{array}{l}
\nabla\.\tilde{\#E}(\#r)=0\\[5pt]
\nabla\.\tilde{\#H}(\#r)=0\\[5pt]
\nabla\times\tilde{\#E}(\#r)=i\omega\muo\mur\tilde{\#H}(\#r)
\\[5pt]
\nabla\times\tilde{\#H}(\#r)=-i\omega\epso\epsr\tilde{\#E}(\#r) 
\end{array}
\right\} 
\end{equation}
in source-free regions.

Plane-wave solutions of Eqs.~\r{diffeq-ehint} are straightforward to find \cite{Chen,BW}.
If the direction of propagation is aligned with the unit vector $\ua$, then
\begin{equation}
\left.\begin{array}{l}
\tilde{\#E}(\#r)= \left(A_{\rm b} \,\ub + A_{\rm c}\,\uc\right)
\exp\left(ik\ua\.\#r\right)
\\[5pt]
\tilde{\#H}(\#r)=\left({\etao\etar}\right)^{-1} \left(A_{\rm b}\, \uc - A_{\rm c}\,\ub\right)
\exp\left(ik\ua\.\#r\right)
\end{array}\right\}\,,
\label{eh-rect}
\end{equation}
where 
$\etar=\sqrt{\mur/\epsr}$ is the  relative impedance,
$k=\ko\sqrt{\mur\epsr}$ is the wavenumber, $A_{\rm b}$ and $A_{\rm c}$ are
amplitudes, the unit vector $\ub\perp\ua$, and the unit vector $\uc=\ua\times\ub$.
Equations~\r{ehint-def} then yield 
\begin{equation}
\left.\begin{array}{l}
{\#E}(\#r)= \left(A_{\rm b} \,\ub + A_{\rm c}\,\uc\right)
\exp\les{ i\left(k\ua+\omega\bGamma\right)\.\#r}\ris
\\[5pt]
{\#H}(\#r)=\left({\etao\etar}\right)^{-1} \left(A_{\rm b}\, \uc - A_{\rm c}\,\ub\right)
\exp\les{ i\left(k\ua+\omega\bGamma\right)\.\#r}\ris
\end{array}\right\}\,
\label{EH-rect}
\end{equation}
for rectilinear propagation.

Whereas Eqs.~\r{eh-rect} describe transverse-electromagnetic plane waves, 
Eqs.~\r{EH-rect} generally do not. This can be seen by decomposing $\bGamma$ into components
parallel and perpendicular to $\ua$ as follows:
\begin{equation}
\bGamma = \left(\bGamma\.\ua\right)\ua + \bGamma\.(\=I-\ua\ua)
\,.
\end{equation} 
Then,
\begin{equation}
\exp\les{ i\left(k\ua+\omega\bGamma\right)\.\#r}\ris=
\exp\les{ i\left(k+\omega\bGamma\.\ua\right)\left(\ua\.\#r\right)}\ris
\exp\les{ i \omega\bGamma\.(\=I-\ua\ua)\.\#r}\ris\,,
\end{equation}
thereby indicating spatial variation in directions perpendicular to $\ua$ in addition
to spatial variation in the direction parallel to $\ua$.

The medium is lossless if (i) $\epsr^\dprime=0$, (ii) $\mur^\dprime=0$, and
(iii) $\bGamma^\dprime=\#0$.
The medium is dissipative if  (i) $\epsr^\dprime>0$, (ii) $\mur^\dprime>0$,
and (iii) $0<\co\vert\bGamma^\dprime\vert<\sqrt{\epsr^\dprime\mur^\dprime}$  \cite{ML2010book}.
Thus, while $\Gamma=\vert\bGamma\vert\ne0$ is responsible
for Lorentz nonreciprocity,
$\bGamma^\prime$ adds  phase and
$\bGamma^\dprime$ adds attenuation.

\section{Reflection and refraction by a half space}\label{randr}
\subsection{Theory}
Suppose that   the  half space $z <0$ is vacuous
but the half space $z>0$ is occupied by a homogeneous medium
with the constitutive relations \r{conrels-EH}.

A plane
wave, propagating in the half space
$z < 0$ at an angle $\thetao\in[0,\pi/2)$ with respect to the $z$ axis and  an angle $\psio\in[0,2\pi)$
with respect to the $x$ axis in the $xy$ plane, is incident on the   plane $z=0$. Thus,
 the  wave vector of the incident plane wave can be stated as
\begin{equation}
\#k\inc=\ko\les(\ux\cpsio+\uy\spsio)\sthetao+\uz\cthetao\ris\,.
\end{equation}
The electromagnetic field phasors associated
with the incident plane wave are represented as
\begin{equation}
\label{EHinc}
\left.\begin{array}{l}
\Einc= (\as\,\sp +\ap \,\pinc) \, \exp\left( i\#k\inc\.\#r\right)
\\[5pt]
\Hinc= \etao  ^{-1} (\as\,\pinc -\ap\, \sp) \, \exp\left( i\#k\inc\.\#r\right)
\end{array}\right\}
\, , \qquad z < 0
\, ,
\end{equation}
 where the  unit vectors
\begin{equation}
\left.\begin{array}{l}
\sp=-\ux\spsio + \uy \cpsio
\\[5pt]
{\#p}_\pm=\mp\le \ux \cpsio + \uy \spsio \ri \cthetao  + \uz \sthetao
\end{array}\right\}
\, 
\end{equation}
identify linear-polarization states.
The amplitudes of the $s$- and the
$p$-polarized components of the incident plane wave, denoted by
$\as$ and
$\ap$, respectively, are assumed given.

The reflected electromagnetic field phasors are expressed as
\begin{equation}
\label{EHref}
\left.\begin{array}{l}
\Erefl= (\rs\,\sp +\rp \,\pref) \, \exp\left(i\#k\refl\.\#r\right)
\\[5pt]
\Hrefl=\etao ^{-1} (\rs\,\pref -\rp\, \sp) \,  \exp\left(i\#k\refl\.\#r\right)
\end{array}\right\}
\, , \qquad z < 0
\, ,
\end{equation}
where the wave vector of the reflected plane wave is
\begin{equation}
\#k\refl=\ko\les(\ux\cpsio+\uy\spsio)\sthetao-\uz\cthetao\ris\,.
\end{equation}
The reflection amplitudes $\rs$ and $\rp$ have to be determined 
in terms of the incidence amplitudes $\as$ and $\ap$ by the solution of
a boundary-value problem.

In order to represent the refracted field phasors, we resort to Eqs.~\r{diffeq-ehint}
to obtain
\begin{equation}
\left.\begin{array}{l}
\tErefr=
\left(
\rhos\,
\spint +\rhop\,
\ppint \right)
\, \exp\les{ i \le \qx x  + \qy y+\alpha z\ri}\ris
\\[5pt]
\tHrefr=\left({\etao\etar}\right)^{-1}
\left(\rhos\,
\ppint
-\rhop\,
\spint \right)
\, \exp\les{ i \le \qx x  + \qy y+\alpha z\ri}\ris
\end{array}\right\}\,,\quad
z>0\,,
 \label{ehint}
\end{equation}
wherein the refraction amplitudes
$\rhos$ and $\rhop$  
are unknown,  
the polarization-state vectors
\begin{equation}
\left.\begin{array}{l}
\spint =\displaystyle{\frac{-\ux\,\qy + \uy \,\qx}{q}}
\\[9pt]
\btau_\pm=
\displaystyle{\mp\le \frac{\ux\, \qx + \uy\, \qy}{q} \ri \frac{\alpha}{k}  + \uz \frac{q}{k}}
\end{array}\right\}
\, ,
\end{equation}
and
the wavenumbers 
$q=+\sqrt{\qx^2+\qy^2}$
and
$\alpha=+\sqrt{k^2-q^2}$
with ${\rm Im}(\alpha)\geq0$.
By virtue of Eqs.~\r{ehint-def}, we  get
\begin{equation}
\left.\begin{array}{l}
\Erefr=
\left(
\rhos\,
\spint +\rhop\,
\ppint \right)
\, \exp\les{ i \le \qx x  + \qy y+\alpha z+\omega\bGamma\.\#r\ri}\ris
\\[5pt]
\Hrefr=\left({\etao\etar}\right)^{-1}
\left(\rhos\,
\ppint
-\rhop\,
\spint \right)
\, \exp\les{ i \le \qx x  + \qy y+\alpha z+\omega\bGamma\.\#r\ri}\ris
\end{array}\right\}\,,\quad
z>0\,.
 \label{EHint}
\end{equation}

The standard boundary conditions across the interface $z=0$ can be satisfied only if
the tangential components of the electric and magnetic field phasors are matched in phase
across the plane $z=0$; accordingly,
\begin{equation}
\left.\begin{array}{l}
\qx=\ko\left(\sthetao\cpsio-\co\Gammax\right)
\\[5pt]
\qy=\ko\left(\sthetao\spsio-\co\Gammay\right)
\end{array}\right\}
\, .
\label{qxqy-def}
\end{equation}
Thus, the wave vector of the refracted plane wave is given by
\begin{equation}
\#k\refr=\#k\inc+\left(\alpha+\omega\Gammaz-\ko\cthetao\right)\uz\,.
\label{krefr-def}
\end{equation}
This equation indicates that
the magnetoelectric-gyrotopic vector $\bGamma$
affects the bending of light upon refraction in two ways. The first is directly
through the presence of $\Gammaz$ in the difference ${\rm Re}\left(\#k\refr\right)-\#k\inc$. The
second is indirectly through the presence of $\Gammax$ and $\Gammay$
in
\begin{equation}
\alpha^2=k^2- \ko^2\les\left(\sthetao\cpsio-\co\Gammax\right)^2+
\left(\sthetao\spsio-\co\Gammay\right)^2\ris\,.
\end{equation}

In order to appreciate the  two ways,
suppose first that $\epsr=\mur=1$ and $\Gamma=0$, i.e., the refracting medium
is also free space.
Reflection is then impossible and the refracted plane wave is the 
same as the incident plane wave.
Next, suppose that $\epsr=\mur=1$ and $\bGamma^\dprime = \#0$ but $\Gamma\ne0$,
so that the refracting medium is the simplest Lorentz-nonreciprocal medium
\cite{Alkhoori2}. Then
the time-averaged Poynting
vector and the phase velocity of the refracted plane wave are not parallel to each other; in other
words,
the ray vector 
\begin{equation}
\#u\refr =\ko^{-1}\#k\inc
\end{equation}
of the refracted wave is not collinear with  $\#k\refr$. Additionally,  the $z$-directed
components of $\#k\refr$ and $\#k\inc$ differ.

Enforcement of the continuities of the  tangential components of the electric and magnetic field phasors  
across the plane $z=0$ yield the solutions
\begin{equation}
\label{r-rho}
\left.\begin{array}{ll}
\rs = \rss\,\as+\rsp\,\ap\,,\qquad & \rhos = \rhoss\,\as+\rhosp\,\ap
\\[5pt]
\rp = \rps\,\as+\rpp\,\ap\,,\qquad & \rhop = \rhops\,\as+\rhopp\,\ap
\end{array}\right\}\,,
\end{equation}
where the products
\begin{eqnarray}
\nonumber
\Delta\cdot\rss &=&-(1-\etar^2) k q^2\alpha\cthetao
-\etar \les(k^2 -\alpha^2\cos^2\thetao)(\qx\spsio-\qy\cpsio)^2
\right.
\\[5pt]
&&\quad\left.
+(\alpha^2-k^2\cos^2\thetao)(\qx\cpsio+\qy\spsio)^2\ris\,,
\label{eq17}
\\[5pt]
\label{rps-def}
\nonumber
\Delta\cdot\rps &=&-\Delta\cdot\rsp
\\[5pt]
&=&2\etar{q^2} (\qx\spsio-\qy\cpsio) (\qx\cpsio+\qy\spsio)\cthetao\,,
\\[5pt]
\nonumber
\Delta\cdot\rpp &=&(1-\etar^2) k q^2\alpha\cthetao
-\etar \les(k^2 -\alpha^2\cos^2\thetao)(\qx\spsio-\qy\cpsio)^2
\right.
\\[5pt]
&&\quad\left.
+(\alpha^2-k^2\cos^2\thetao)(\qx\cpsio+\qy\spsio)^2\ris\,,
\\[5pt]
\Delta\cdot\rhoss &=&
2\etar{kq}
(\alpha\etar+k\cthetao)(\qx\cpsio+\qy\spsio)\cthetao\,,
\\[5pt]
\label{rhops-def}
\Delta\cdot\rhops &=&
2\etar{kq}
(k\etar+\alpha\cthetao)(\qx\spsio-\qy\cpsio)\cthetao\,,
\\[5pt]
\label{rhosp-def}
\Delta\cdot\rhosp &=&
-2\etar{kq}
(k +\alpha\etar\cthetao)(\qx\spsio-\qy\cpsio)\cthetao\,,
\\[5pt]
\Delta\cdot\rhopp &=&
2\etar{kq}
(\alpha +k\etar\cthetao)(\qx\cpsio+\qy\spsio)\cthetao\,,
\end{eqnarray}
involve
\begin{eqnarray}
\nonumber
\Delta &=&(1+\etar^2) k q^2\alpha\cthetao
+\etar \les(k^2 +\alpha^2\cos^2\thetao)(\qx\spsio-\qy\cpsio)^2
\right.
\\[5pt]
&&\quad\left.
+(\alpha^2+k^2\cos^2\thetao)(\qx\cpsio+\qy\spsio)^2\ris\,.
\label{eq24}
\end{eqnarray}
The foregoing expressions reduce to  standard results \cite{Iskander}
when $\Gammax=\Gammay=0$. If the refracting medium is matched in 
impedance to free space, i.e., $\etar=1$,
Eqs.~\r{eq17}--\r{eq24} yield: $\rss=\rpp$, $\rhoss=\rhopp$, and $\rhops=-\rhosp$.

The four reflection coefficients in Eqs.~\r{r-rho}
are denoted by ${r_{\rm ab}}$  and the four
refraction coefficients by ${\rho_{\rm ab}}$, 
$ a\in\lec{p,s}\ric$ and 
$b\in\lec{p,s}\ric$. Co-polarized coefficients have both
subscripts identical, but cross-polarized coefficients do not. Four reflectances
are defined as ${R_{\rm ab}}=\vert{r_{\rm ab}}\vert^2$. The principle of conservation of
energy requires that
\begin{equation}
\left.\begin{array}{l}
0\leq\Rss+\Rps \leq 1
\\[5pt]
0\leq\Rpp+\Rsp\leq1
\end{array}
\right\}\,.
\end{equation}
The difference $1-\left(\Rss+\Rps\right)$ 
is the fraction of the incident energy that is refracted into the half space
$z>0$ if the incident plane wave is $s$ polarized, and
 $1-\left(\Rpp+\Rsp\right)$ is the analogous quantity if the incident plane wave is $p$ polarized.

\subsection{Numerical results and discussion}
Whereas $\Gammaz$ does not appear in Eqs.~\r{eq17}--\r{eq24} because
the bi-medium interface is the plane $z=0$, both $\Gammax$ and $\Gammay$ do. One way to quantify
their effect is via  
\begin{equation}
\psi_q=\tan^{-1}\left(\qy/\qx\right)\,,
\end{equation}
which is a real angle if
$\Gammax^\dprime=\Gammay^\dprime=0$. In that case, $\psi_q\ne\psi$ unless
$\Gammax^\prime=\Gammay^\prime=0$ also. Thus, the magnetoelectric-gyrotropy vector
can rotate the refracted light about the $z$ axis even if the refracting medium is
the simplest Lorentz-nonreciprocal medium  \cite{Alkhoori2}.

\begin{figure}[h]
 \centering 
\includegraphics[width=0.6\linewidth]{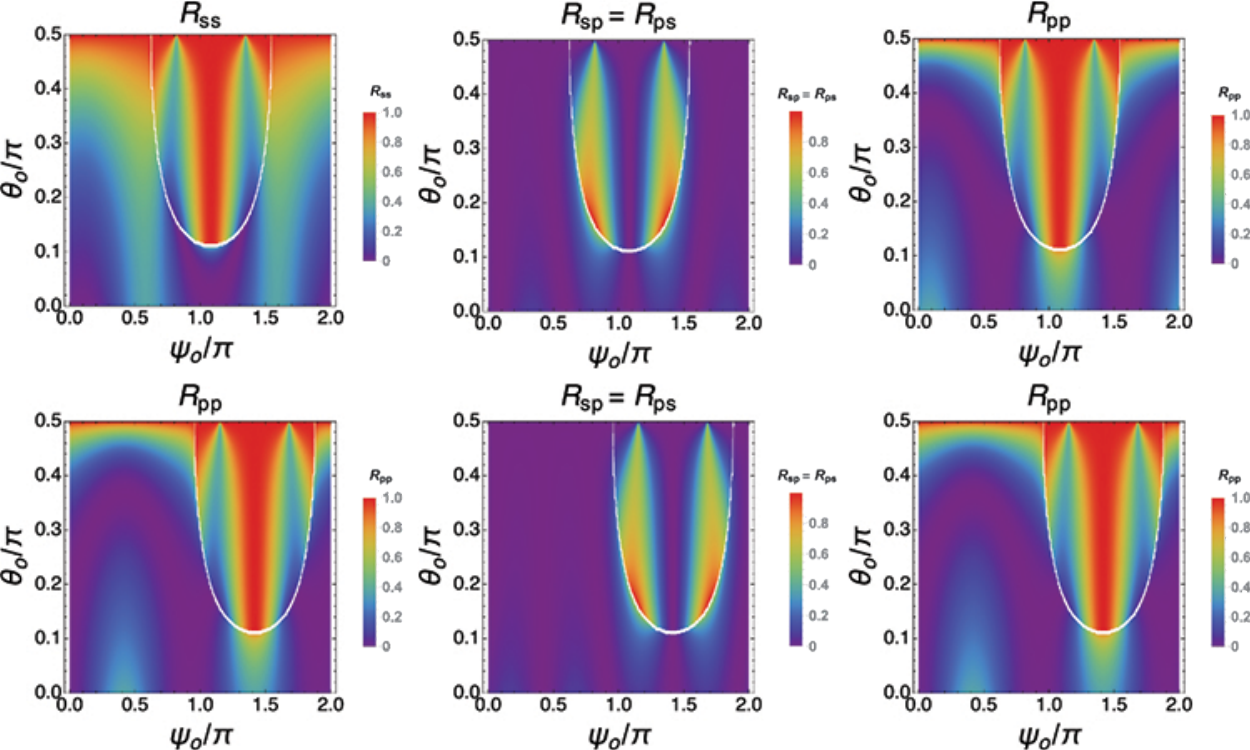}
\caption{Reflectances  
as functions of the incidence angles $\thetao$ and $\psio$,
when a plane wave propagating in free
space is reflected and refracted by a half space occupied by
a Lorentz-nonreciprocal medium described by
Eqs.~\r{conrels-EH} with $\epsr=5$, $\mur=1.1$, and $\co \bGamma= 2(\ux\cos\varphi+\uy\sin\varphi)$.
The  reflectances are unaffected by $\Gammaz$.
Top row: $\varphi=15^\circ$. Bottom row: $\varphi=75^\circ$. The white line around the bubble
in each plot is a numerical artifact arising from the discrete values of $\thetao$ and $\psio$
used to generate the plot.
}
\label{Figure1}
\end{figure}

Figure~\ref{Figure1} shows the variations of the four reflectances $R_{\rm ab}$ with $\thetao$ and $\psio$, when
$\epsr=5$, $\mur=1.1$, and $\co\bGamma=2(\ux\cos\varphi+\uy\sin\varphi)$.
Data are presented for $\varphi\in\lec15^\circ,75^\circ\ric$. For any fixed $\thetao$,
the reflectances calculated for $\varphi=75^\circ$ are shifted on the $\psio$ axis
by $60^\circ$ in relation to the reflectances for $\varphi=15^\circ$. Equations~\r{eq17}--\r{eq24}
predict this shift because all reflection and refraction
coefficients are functions of $\psio-\varphi$, but not of $\psio$ and $\varphi$ separately,
when $\bGamma$ can be expressed in the form $\Gamma_{\rm xy}(\ux\cos\varphi+\uy\sin\varphi)
+\Gammaz\uz$, even if $\Gamma_{\rm xy}$ is complex.

\begin{figure}[h]
 \centering 
\includegraphics[width=0.6\linewidth]{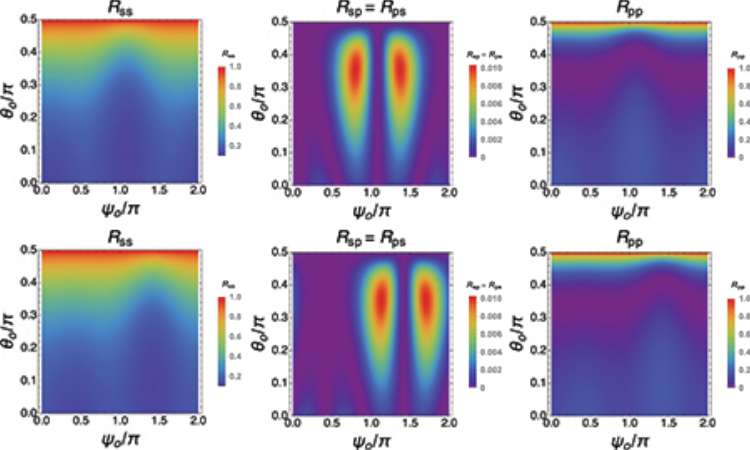}
\caption{Same as Fig.~\ref{Figure1} except that
$\co\bGamma=0.9(\ux\cos\varphi+\uy\sin\varphi)$. 
}
\label{Figure2}
\end{figure}

The prominent bulbous features in Fig.~\ref{Figure1} arise because $\alpha$ changes from purely real
to purely imaginary and then to purely real again, when $\thetao$ is kept fixed but $\psio$ is 
varied continuously. These features do not exist if $\Gamma_{\rm xy}$ is reduced to
a sufficiently small and real value,
as is exemplified by the reflectance plots in Fig.~\ref{Figure2} for $\co\Gamma_{\rm xy}=0.9$.
The bulbous features also vanish when the refracting medium is dissipative.

Both $\Gammax$ and $\Gammay$, singly and jointly, lead to depolarization upon reflection and refraction.
Expressions for $\rps=-\rsp$, $\rhops$, and $\rhosp$ are directly proportional to
the term $(\qx\spsio-\qy\cpsio)$, as is clear from
Eqs.~\r{rps-def}, \r{rhops-def}, and \r{rhosp-def}. But $\qx\spsio\to\qy\cpsio$
as $\Gammax\to0$ and $\Gammay\to0$, according to Eqs.~\r{qxqy-def}.

Also, both $\Gammax$ and $\Gammay$, singly and jointly, cause 
asymmetry in reflection with respect to the reversal of projection of the propagation
direction of the incident plane wave on the illuminated plane $z=0$. The phenomenon
of left/right reflection asymmetry \cite{LMjnp2016} can be quantitated
through
\begin{equation}
\Delta{R_{\rm ab}}(\thetao,\psio)= {R_{\rm ab}}(\thetao,\psio)-{R_{\rm ab}}(\thetao,\psio+\pi)\,,
\quad
 a\in\lec{p,s}\ric\,,\quad
 b\in\lec{p,s}\ric\,.
\end{equation}
Left/right reflection asymmetry exists if $\Delta{R_{\rm ab}}(\thetao,\psio)\ne0$ for at least one
of the four
combination of the subscripts $a$ and $b$.

\begin{figure}[h]
 \centering 
\includegraphics[width=0.5\linewidth]{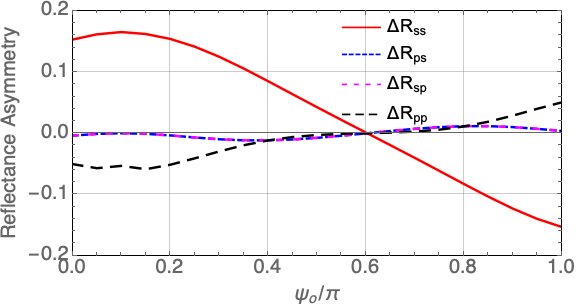}
\caption{Left/right reflection asymmetry in relation to $\psio\in[0,\pi]$
when a plane wave propagating in free
space is reflected and refracted by a half space occupied by
a Lorentz-nonreciprocal medium described by
Eqs.~\r{conrels-EH} with $\epsr=5+i0.4$, $\mur=1.1+i0.05$,   $\co\Gammax=0.9+i0.01$,
and
$\co\Gammay=0.3+i0.01$; $\thetao=\pi/3$.
}
\label{Figure3}
\end{figure}

Although left/right reflection asymmetry  can be visually inferred from Figs.~\ref{Figure1}
and \ref{Figure2}, it can be grasped more quickly
from the plots of $\Delta{R_{\rm ab}}(\pi/3,\psio)$ 
versus $\psio\in[0,\pi]$ in Fig.~\ref{Figure3} when the refracting medium is taken to be
dissipative.
Left/right asymmetry in cross-polarized reflectances is considerably
weaker than in co-polarized
reflectances. Furthermore,
the degrees of left/right asymmetry of $\Rss$ and $\Rpp$ can be quite different. Thus,
the simple Lorentz-nonreciprocal medium with magnetoelectric gyrotropy is promising
for   one-way optical devices that could be used to reduce backscattering noise as well as 
instabilities in  communication networks and   various imaging devices for
 microscopy and tomography \cite{Jalas,Sayrin}.

\section{Reflection by and transmission through a slab}\label{randt}
\subsection{Theory}
Suppose that   the  half spaces $z <0$ and  $z>L$ are vacuous
but the slab $0<z<L$ is composed of a homogeneous medium
with the constitutive relations \r{conrels-EH}. 

As in Sec.~\ref{randr},
a plane
wave, propagating in the half space
$z < 0$ at an angle $\thetao\in[0,\pi/2)$ with respect to the $z$ axis and at an angle $\psio\in[0,2\pi)$
with respect to the $x$ axis in the $xy$ plane, is incident on the   plane $z=0$. Equations~\r{EHinc}
and \r{EHref} still apply in the half space $z<0$ with unknown reflection amplitudes
$\rs$ and $\rp$, but the incidence amplitudes $\as$ and $\ap$
are presumed known.
The  wave vector of the  plane wave transmitted into the half space $z>L$
is  $\#k\inc$, whereas the associated electromagnetic field phasors
can be stated as
\begin{equation}
\label{EHtr}
\left.\begin{array}{l}
\Etr= \left(\ts\,\sp +\tp\, \pinc\right) \, 
\exp\left[ i\#k\inc\.\left(\#r-L\uz\right)\right]
\\[5pt]
\Htr=  \etao ^{-1}\left(\ts\,\pinc -\tp\, \sp\right) 
\exp\left[ i\#k\inc\.\left(\#r-L\uz\right)\right]
\end{array}\right\}
\, , \qquad z > L
\, ,
\end{equation}
where the transmission amplitudes $\ts$ and $\tp$ are unknown.

A standard $4\times4$-matrix procedure to solve the boundary-value problem yields
\cite{LWprsa2}  
\begin{equation}
\left[\begin{array}{l} t_s\\t_p\\ 0 \\0\end{array}\right]=
[\=K]^{-1}\.
\exp\left\{{i[\=P]L}\right\}\.[\=K]\.
\left[\begin{array}{l} a_s\\a_p\\ r_s \\r_p\end{array}\right]\,,
\label{finaleq}
\end{equation}
where the $4\times4$ matrixes
\begin{equation}
[\=K] =
\left[ \begin{array}{cccc}
-\spsio & -\cpsio\cthetao  & -\spsio & \cpsio\cthetao 
\\[5pt]
\cpsio & -\spsio\cthetao  & \cpsio & \spsio\cthetao 
\\[5pt]
- \etao^{-1}\cpsio\cthetao & 
\etao^{-1}\spsio  
&\etao^{-1}\cpsio\cthetao &
\etao^{-1}\spsio 
\\ [5pt]
-\etao^{-1}\spsio\cthetao &
-\etao^{-1}\cpsio &
\etao^{-1}\spsio\cthetao 
& -\etao^{-1}\cpsio
\end{array}\right]\,
\end{equation}
and
\begin{equation}
[\=P]=
\displaystyle{\left[\begin{array}{cccc}
\omega\Gammaz & 0 & \frac{\qx\qy}{\omega\epso\epsr} &\omega\muo\mur- \frac{\qx^2}{\omega\epso\epsr}
\\[5pt]
0 & \omega\Gammaz &-\omega\muo\mur+ \frac{\qy^2}{\omega\epso\epsr} &
-\frac{\qx\qy}{\omega\epso\epsr}
\\[5pt]
-\frac{\qx\qy}{\omega\muo\mur} & -\omega\epso\epsr+ \frac{\qx^2}{\omega\muo\mur} 
&\omega\Gammaz & 0 
\\[5pt]
\omega\epso\epsr- \frac{\qy^2}{\omega\muo\mur} &\frac{\qx\qy}{\omega\muo\mur} 
&0 & \omega\Gammaz
\end{array}\right]}\,.
\end{equation}

Equation~\r{finaleq} can be solved by numerical means \cite{Jaluria,Chapra} to determine
the four reflection coefficients  ${r_{\rm ab}}$ and the four transmission coefficients
 ${t_{\rm ab}}$, $ a\in\lec{p,s}\ric$ and 
$b\in\lec{p,s}\ric$, that appear in the following relations:
\begin{equation}
\label{r-t}
\left.\begin{array}{ll}
\rs = \rss\,\as+\rsp\,\ap\,,\qquad & \ts = \tss\,\as+\tsp\,\ap
\\[5pt]
\rp = \rps\,\as+\rpp\,\ap\,,\qquad & \tp = \tps\,\as+\tpp\,\ap
\end{array}\right\}\,.
\end{equation}
Four reflectances
are defined as $\Rsp=\vert\rsp\vert^2$, etc., and four transmittances
as $\Tsp=\vert\tsp\vert^2$, etc.
The principle of conservation of
energy requires that
\begin{equation}
\left.\begin{array}{l}
0\leq\Rss+\Rps +\Tss+\Tps \leq 1
\\[5pt]
0\leq\Rpp+\Rsp+\Tpp+\Tsp\leq1
\end{array}
\right\}\,.
\end{equation}
The differences $1-\left(\Rss+\Rps+\Tss+\Tps\right)$ and $1-\left(\Rpp+\Rsp+\Tpp+\Tsp\right)$
indicate the fraction of the incident energy that is absorbed in the slab $0<z<L$.

A word of caution: The reflection coefficients appearing in Eqs.~\r{r-t} are not the same
as the reflection coefficients appearing in Eqs.~\r{r-rho}. The obvious difference
between the two sets of reflection coefficients is that the ones in Eqs.~\r{r-t} depend
on $L$ and $\Gammaz$ but the ones in Eqs.~\r{r-rho} do not.

\subsection{Numerical results and discussion}

\begin{figure}[h]
 \centering 
\includegraphics[width=0.45\linewidth]{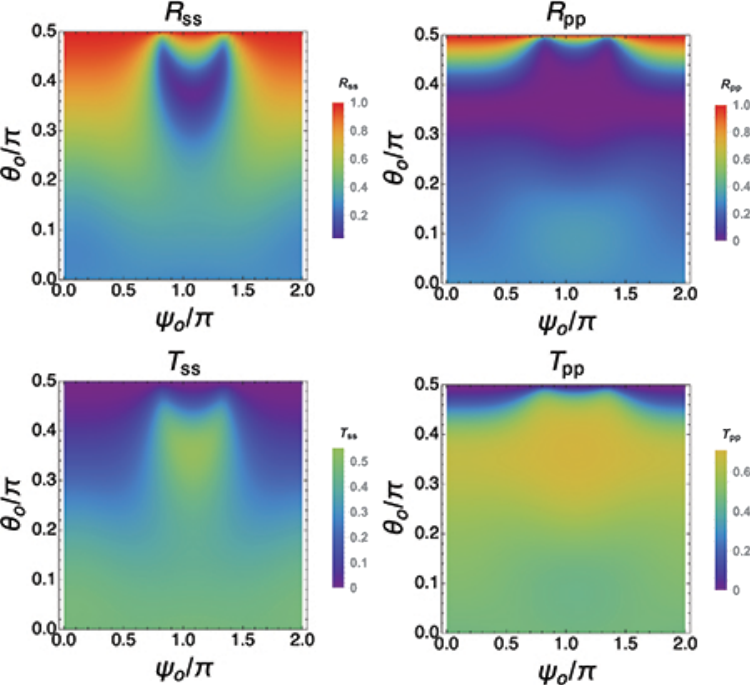}
\caption{Co-polarized reflectances and transmittances 
as functions of the incidence angles $\thetao$ and $\psio$,
when a plane wave propagating in free
space is incident on a 1-wavelength-thick slab (i.e., $L=\lambdao$) composed of
a Lorentz-nonreciprocal medium described by
Eqs.~\r{conrels-EH} with $\epsr=5+i0.04$, $\mur=1.1+i0.01$, and $\co\bGamma=(0.3+i0.01)(\ux\cos\varphi+\uy\sin\varphi)+(0.5+i0.01)\uz$ with $\varphi=15^\circ$. The cross-polarized
reflectances and transmittances are $<10^{-3}$ in magnitude and therefore are not presented.
}
\label{Figure4}
\end{figure}

\begin{figure}[h]
 \centering 
\includegraphics[width=0.45\linewidth]{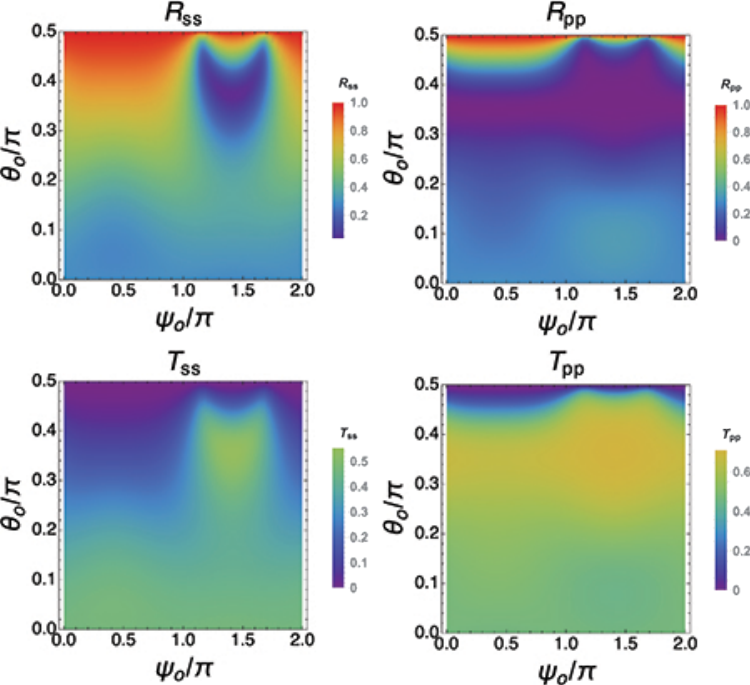}
\caption{Same as Fig.~\ref{Figure4}, except that  $\varphi=75^\circ$.
}
\label{Figure5}
\end{figure}

Figure~\ref{Figure4} shows the co-polarized reflectances and transmittances
calculated as functions of the incidence angles $\thetao$ and $\psio$,
when $L=\lambdao$, $\epsr=5+i0.04$, $\mur=1.1+i0.01$, and $\co\bGamma=(0.3+i0.01)(\ux\cos\varphi+\uy\sin\varphi)+(0.5+i0.01)\uz$ with $\varphi=15^\circ$. The cross-polarized
reflectances and transmittances are $<10^{-3}$ in magnitude and therefore are not shown.
Calculations for
Fig.~\ref{Figure5} were made with the same parameters as for Fig.~\ref{Figure4}, except that $\varphi$ was increased to $75^\circ$. A comparison of the two figures quickly reveals
that,
for any fixed $\thetao$,
all remittances calculated for $\varphi=75^\circ$ are shifted on the $\psio$ axis
by almost $60^\circ$ in relation to their counterparts calculated for $\varphi=15^\circ$. Thus, the
ability of the transverse component $\left(\bGamma-\Gammaz\uz\right)$ of the magnetoelectric-gyrotropy
vector $\bGamma$
to cause rotation about the $z$ axis is evinced, just as in Sec.~\ref{randr},  
for reflection and transmission by a slab.

\begin{figure}[h]
 \centering 
\includegraphics[width=0.45\linewidth]{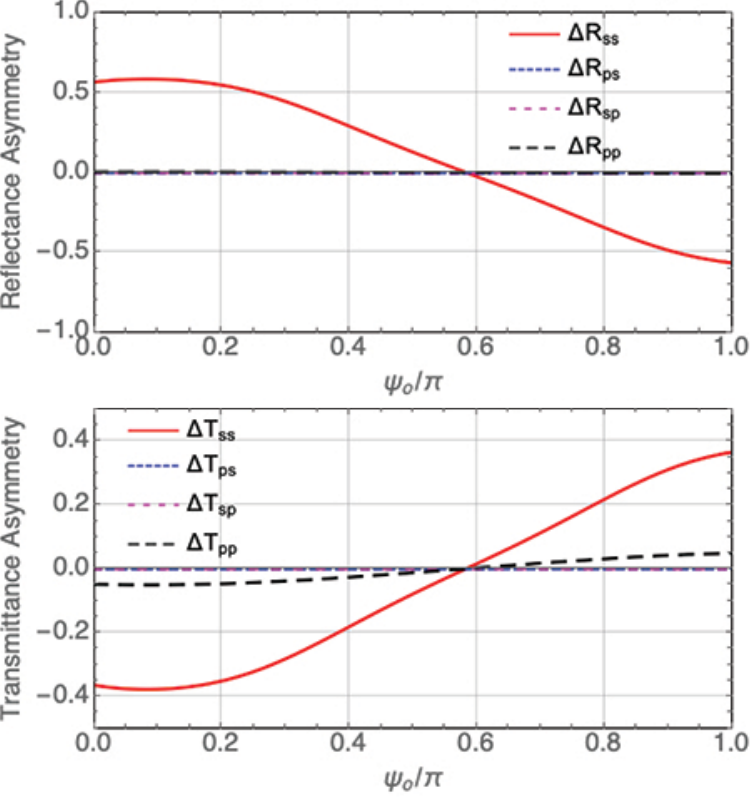}
\caption{Left/right reflection asymmetry and transmission
asymmetry in relation to $\psio\in[0,\pi]$
when a plane wave propagating in free
space is  incident on a 1-wavelength-thick slab (i.e., $L=\lambdao$)
composed of
a Lorentz-nonreciprocal medium described by
Eqs.~\r{conrels-EH} with $\epsr=5+i0.04$, $\mur=1.1+i0.01$, and $\co\bGamma=(0.3+i0.01)(\ux\cos\varphi+\uy\sin\varphi)+(0.5+i0.01)\uz$ with $\varphi=15^\circ$; $\thetao=\pi/3$.
}
\label{Figure6}
\end{figure}

Left/right asymmetry is evident for the reflectances in Figs.~\ref{Figure4} and \ref{Figure5},
just as in Sec.~\ref{randr}. Left/right asymmetry is evident 
also for the transmittances in both figures.
The degree of left/right transmission asymmetry, as quantified by
\begin{equation}
\Delta{T_{\rm ab}}(\thetao,\psio)= {T_{\rm ab}}(\thetao,\psio)-{T_{\rm ab}}(\thetao,\psio+\pi)\,,
\quad
 a\in\lec{p,s}\ric\,,\quad
 b\in\lec{p,s}\ric\,,
\end{equation}
is comparable to the degree of left/right reflection asymmetry. This becomes clear from
the plots of $\Delta{R_{\rm ab}}(\pi/3,\psio)$ and $\Delta{T_{\rm ab}}(\pi/3,\psio)$ presented
in Fig.~\ref{Figure6}.

\begin{figure}[h]
 \centering 
\includegraphics[width=0.45\linewidth]{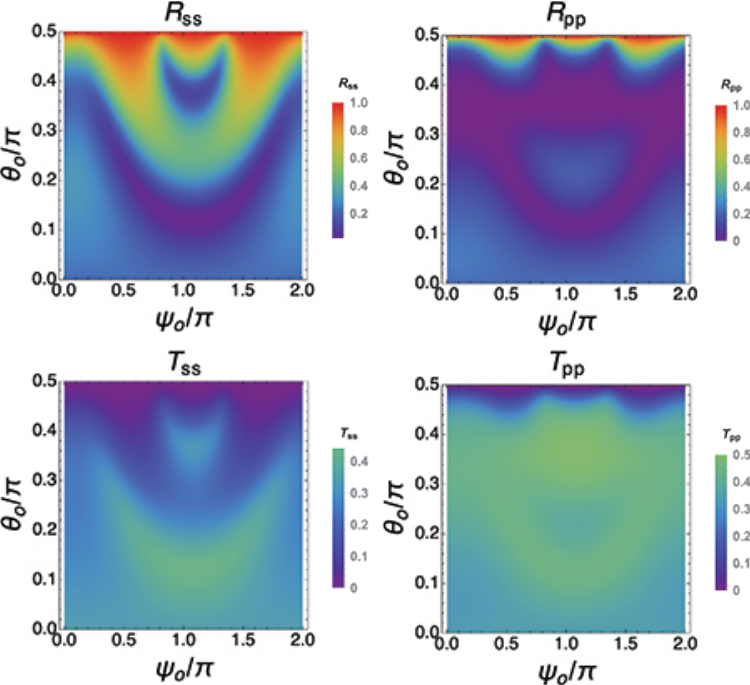}
\caption{Same as Fig.~\ref{Figure4}, except that  $L=2\lambdao$.
}
\label{Figure7}
\end{figure}

The effect of $\Gammaz$ can be appreciated when the Lorentz-nonreciprocal medium with
magnetoelectric gyrotropy  
occupies a finite extent along the $z$ axis, as in this section. The plots in
Fig.~\ref{Figure7} were obtained with the same parameters as those in Fig.~\ref{Figure4},
 except that $L$ was increased from $\lambdao$ to $2\lambdao$, so that the slab for the former figure is twice as thick
as for the latter figure.  The transmittances  are smaller for the higher value of $L/\lambdao$,
and so are the reflectances in general. Let us recall, however, that all reflectances, transmittances,
and absorptances have an upper limit of unity.

\section{Concluding remarks}
The Pl\'ebanski theorem of general relativity inspires a bianisotropic medium with:
(i) isotropic permittivity, (ii) isotropic permeability, and (iii) magnetoelectric
gyrotropy. This  is a simple Lorentz-nonreciprocal medium that can be transformed
into an isotropic dielectric-magnetic medium. Accordingly,
field representations in the Lorentz-nonreciprocal medium
 can be obtained from those for
the isotropic dielectric-magnetic medium. 

When a plane wave is incident on a half space occupied
by the  Lorentz-nonreciprocal medium with magnetoelectric gyrotopy, theory shows that
the transverse component of the magnetoelectric-gyrotopy vector is responsible
for a rotation about the normal axis. Furthermore, left/right reflection
asymmetry is exhibited. A planar slab of the Lorentz-nonreciprocal medium with magnetoelectric gyrotopy
exhibits
left/right transmission asymmetry in addition to left/right reflection asymmetry. Thus, magnetoelectric
gyrotropy is a constitutive mechanism that delivers left/right asymmetry in both reflection and transmission,
a phenomenon  of major interest to realize
one-way  devices in optics.

\vspace{1cm}

\noindent \textbf{Acknowledgment.}
The author is grateful to the Charles Godfrey
Binder Endowment at Penn State for ongoing support of his research.

\vspace{1cm}

\noindent {\bf Declarations
of interest}: none

\end{document}